\documentclass[preprint,]{elsarticle}
\usepackage{graphicx, amssymb}
\begin{document}

\begin{frontmatter}
\title{Simulation of ion track ranges in uranium oxide}
\author[das]{Byoungseon Jeon \corref{cor}}
\ead{bjeon@ucdavis.edu}
\author[ucb]{Mark Asta}
\author[lanl]{Steven M.~Valone}
\author[das]{and Niels~Gr{\o}nbech-Jensen}
\cortext[cor]{Corresponding author}
\address[das]{Department of Applied Science, University of California, 
Davis, CA 95616}
\address[ucb]{Department of Material Science and Engineering, University of
California, Berkeley, CA 94720, and Department of Chemical Engineering and
Materials Science, University of California, Davis, CA 95616}
\address[lanl]{Materials Science and Technology Division, Los Alamos National Laboratory, Los Alamos, NM 87544}

\date{\today}

\begin{abstract}
Direct comparisons between statistically sound simulations of ion-tracks and
published experimental measurements of range densities of iodine implants in
uranium dioxide have been made with implant energies in the range of 
100-800 keV. Our simulations are conducted with REED-MD (Rare Event Enhanced
Domain-following Molecular Dynamics) in order to account 
for the materials structure in both single crystalline and polycrystalline 
experimental samples. We find near-perfect agreement between REED-MD results 
and experiments for polycrystalline target materials. 
\end{abstract}

\begin{keyword}
REED-MD \sep ion-track range \sep molecular dynamics \sep nuclear fuel materials \sep uranium oxide
\PACS 31.15.xv \sep 07.05.Tp \sep 61.80.-x  
% Molecular dynamics calculations in atomic and molecular physics,
% Computer modeling and simulation, Irradiation effects in solids
% http://www.aip.org/pacs/pacs2010/individuals/pacs2010_regular_edition/index.html
\end{keyword}

\end{frontmatter}

\section{Introduction}
Ion irradiation of materials with subsequent analysis of associated dopant 
density distributions by, e.g., Secondary Ion Mass Spectroscopy (SIMS) is a 
powerful tool for understanding irradiation effects in materials \cite{ZBL}. 
Dopant implantation into crystalline target 
materials provides especially detailed information about effective ionic size, 
atomic force-field, interatomic interactions, particularly at close
distances, and the role of dissipation by electronic excitations.
This is due to the variety of typical interaction mechanisms 
reflected in the statistics of ion trajectories which are influenced by
a variety of processes ranging from 
direct atomic collisions, to propagation through open channels of a given
crystal structure with relatively minor interactions with the surrounding
environment. The analyses of sets of carefully prepared SIMS profiles 
from different selected ion implantation experiments can therefore provide  
a wealth of information and the validation of 
theoretical models, which, in turn, can be used to interpret other 
experimentally generated SIMS profiles.

Simulating dopant density distributions due to ion implantation is 
naturally done with atomic representations, since the evolution of a 
trajectory is given by its atomic scale interactions with the target 
material. However, the computational practicality of conducting just a 
few complete simulations of high-energy ions is unrealistic given
that many trajectories for ions with  MeV energies have ranges of the order 
of micrometers. This is especially true for relevant radiation effects in 
nuclear fuel materials, where fission products have energies in the 100 MeV 
range and $\alpha$-decay of heavy elements results in recoils in the range 
of 100 keV. Thus, the simulated volume of material needs to include billions 
of atoms in a Molecular Dynamics (MD) simulation scheme.
Also due to the very high initial ion energy, the necessary time step of 
a numerical integrator for advancing the simulation is so small that an 
unpractical number of time steps is needed in order to resolve 
the dynamics during high-energy atomic collisions. One additional
obstacle to easy production of simulation data that can be directly compared 
to experimental ion-ranges is that experimental data is a statistical 
composite of a very large number of ion-trajectories that make up the 
resulting measured distributions. Thus, one needs to conduct enough ion 
track simulations 
for a statistically meaningful density profile of resting ions to be obtained.

The conventional Binary Collision Approximation (BCA) addresses at least the 
first two of these problems, namely, the spatial and temporal scales. 
This is done by neglecting all ion-target interactions that are not primary 
to the ion; i.e., the ion only interacts with a single target atom at a time,
and this interaction is modeled as a complete and instantaneous trajectory 
of a collision. While this approximation has proven to be extremely powerful 
for trajectories in materials where crystal structure and channeling
can be neglected, such as very dense materials, disordered materials, 
or for ions that are large compared to the crystal channel dimensions
\cite{eckstein}, 
it has inherent deficiencies when channeling along 
characteristic crystal orientations becomes important.

Here we apply the Rare Event Enhanced Domain following 
Molecular Dynamics (REED-MD) technique, which is a hybrid method combining
elements of the  BCA and MD in order to provide 
physically relevant and computationally practical simulations of ion-implant 
density profiles in crystalline uranium dioxide. This algorithm,
which has been described in detail elsewhere \cite{beardmore98,reed_par},
simulates only a limited amount of the target material at any given time, 
while providing a full-symmetry description of the forces acting on the ion 
by the target material. Thus, tracks along channels are well accounted for.
An additional feature of the algorithm is the ability to obtain near-constant 
statistical uncertainty over many orders of magnitude in the simulated 
density. This increases the simulation efficiency by
as many orders of magnitude as one wishes to resolve the density.

As is well known \cite{Kleykamp85,mattews87,Gittus89} and recently 
re-emphasized \cite{saidy08}, the behavior of 
fission gas products can have a substantial impact on nuclear fuel 
performance and safety.  After their formation during reactor operation, 
these species can diffuse to fuel-cladding interfaces and/or be 
released from the fuel.  Either of these outcomes can have a 
deleterious effect on the reactor. Fission-gas release can also 
occur under various storage and disposal conditions.  For these 
reasons it is valuable to be able to simulate the profiles of fission 
gas products as they are produced.  Here, we apply our simulation methodology 
in comparison with published experiments of 100-800 keV iodine implants 
into UO$_2$ targets of different 
structural characteristics.

\section{Implementation of REED-MD}
The physics implemented into REED-MD for this application consists
of different components. Atomic interactions are modeled by the
universal screened core potential of Ziegler, Biersack, and Littmark (ZBL)
\cite{ZBL}, which objectively represents the effective screening of one
atom in the presence of another. The computationally efficient functional
form for the potential energy $V(r)$ of interaction between two atoms at a 
mutual distance $r$ is given by the core
repulsion between the nuclei along with a simple universal screening function
\begin{equation}
V(r) = {e^2 \over 4 \pi \epsilon_0 r} Z_1 Z_2 \phi_u,
\end{equation}
where $Z_i$ is the atomic number of atom $i$, $\epsilon_0$ is the vacuum
permittivity constant, and $e$ is the unit of charge.
The screening function is defined as
\begin{equation}
\phi_u = 0.18175 e^{-3.1998x} + 0.50986 e^{-0.94229x} + 0.28022 e^{-0.4029x} 
+ 0.028171 e^{-0.20162x},
\label{eqn:screen}
\end{equation}
where $x=r/a_U$ is a reduced pair distance, with $a_U$ being the universal 
screening length
\begin{equation}
a_U = \frac{a_B}{Z_1^{0.23} + Z_2^{0.23}} \; ,
\end{equation}
where $a_B=\sqrt[3]{\frac{9\pi^2}{128}}r_B$, $r_B\approx0.52918${\AA} being 
the Bohr radius. This interaction accounts neither for the electrostatic 
interactions in the ionic material nor for the details of chemical bonding. 
However, these particular interactions are negligible
compared to the collision energy that determines
the range profiles of high-energy ions. Thus, for this particular purpose, 
the simplicity
and objectivity of the universal ZBL interaction is sufficient 
since we do not concern ourselves with the low energy and near-equilibrium
properties of the material.

Since molecular dynamics does not take electronic degrees of freedom (DOF) 
into account in an explicit manner that allows for direct energy transfer
from moving atoms to those DOF,
we adopt two phenomenological mechanisms for energy dissipation.
Firsov's inelastic collisions between atoms as described by 
Elteckov \cite{elteckov}, is expressed as a function of the relative 
velocity between two atoms leading to the following expression for the 
interatomic force:
\begin{equation}
{\bf F}_{ij}(r) =  {0.7 \hbar \over \pi r_B^2}  ( {\bf v}_j - {\bf v}_i ) 
\left[ \frac{Z_i^2}{\left(1+0.8\alpha Z_i^{1/3} r /a_B\right)^4} + 
\frac{ Z_j^2}{ \left(1 + 0.8(1-\alpha) Z_j^{1/3} r/a_B\right)^4}
\right]\; ,
\end{equation}
\begin{equation}
\alpha = \left[ 1 + \left( {Z_j \over  Z_i} \right)^{1/6} \right]^{-1},
\end{equation}
where $Z_i \ge Z_j$. 

Second, electronic stopping is employed using a slightly revised Brandt and 
Kitagawa 
formulation \cite{bk82,cai96}. This phenomenological description connects  
drag on a moving atom with the atom type, the atom velocity, and the electron 
density which it is moving through. Thus, this term acts as local friction 
in the equation of motion for the atom of the form 
\begin{equation}
\Delta E_e = \int_{\rm ion ~path} (Z_1^*)^2 S(v_1, r_s(x)) dx
\label{eqn:stop_pow}
\end{equation}
where $Z_1^*$ is the effective charge of the moving ion, and $S$ is the
stopping power of a unit charge. In the formulation of the stopping power,
we employ an adjustable parameter $r_s^0$, being a mean value of one-electron
radius (Wigner-Seitz radius). This value can be determined empirically and 
allows the fine tuning of range profile. In this study, we used 
$r_s^0 = 1.20 \rm\AA$ for UO$_2$ tests. Detailed description of the 
formulation can be found in the Ref. \cite{bk82,cai96}.

Given that only short range interactions are included in the dynamics of a 
moving ion we adopt a domain following strategy \cite{Nordlund95}, in which 
the ion of interest is modeled along with the unit cell it is currently in,
as well as a buffer of one more unit cell layer surrounding
the ion. For a standard stoichiometric fluorite UO$_2$ structure this 
becomes 325 simulated atoms with the moving ion interacting with target 
material up to a distance longer than one unit cell ($>5${\AA}). 
All these atoms interact, but as the ion is moving out of one unit cell,
and into the next, new unit cells are created ahead of the track, 
while material is discarded in its wake such that the ion always is 
located in the center of the $3\times3\times3$ unit cells.
We thereby simulate the interaction of a moving ion with an 
infinite bulk target material. 
We construct the new unit cell material according to the thermally 
disordered fluorite structure at a given temperature, employing 
a Gaussian distribution of thermal atomic displacements.

Molecular dynamics of an ion track is modeled as prescribed above
and with adaptive time step control according to both the kinetic energy of
the ion and the potential energy of atomic collisions. One such 
simulation is initiated with an incident ion inserted with a random position 
within a unit cell of the target material, 
and an initial momentum according to the energy and direction of the ion track.
The trajectory is then followed until the ion slows to a given threshold 
energy  when the ion is defined to be at rest. 
The distribution of implanted ion depths is obtained by conducting many such 
simulations in order to provide a statistically representative ensemble 
of outcomes. To this end, we dynamically optimize several distances at 
which the system is cloned into the multiplicity of separately evolving
simulations with appropriate statistical weights. Rare event enhancement
ensures that the algorithm consumes nearly the same amount of computing 
resources deriving statistics for each  implant depth, 
even if the physical number of atoms at different 
depths is vastly different. As a result, we can obtain good statistics at 
all scales of interest, which is especially important for fission product
profiles with appreciable densities at large depths due to channeling. These 
two ingredients represent the standard features of the REED-MD algorithm;
more detailed descriptions can be found in Ref. \cite{beardmore98,cai98,beardmore99}. 
Recently REED-MD has been parallelized with standard MPI (Message Passing 
Interface) \cite{reed_par}, and ions of  up to $\sim 100$ MeV  
can be handled with adequate statistics when about $10^4$ initial ions are 
simulated. 

\begin{figure}[p]
    \centering
\includegraphics[clip,scale=0.4, angle=-90]{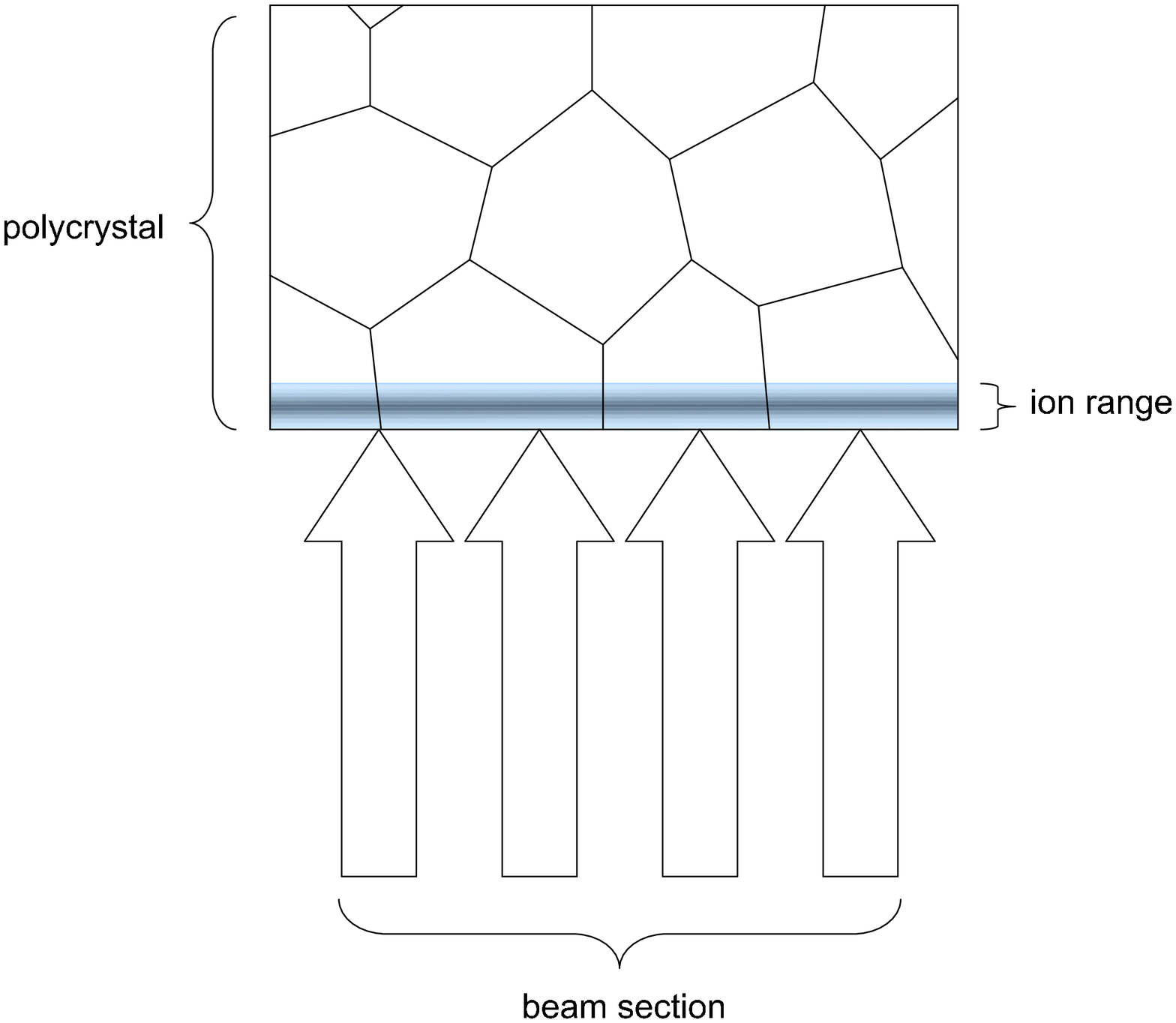}
\caption{Schematic diagram of the geometry associated with the ion implantation experiments for polycrystalline $\rm UO_2$.}
\label{fig:poly}
\end{figure}

\section{Modeling Implant Densities in Polycrystalline Samples}
The structure of the material is one of several important parameters when 
studying ion range density profiles in crystalline targets. The orientation 
of the material relative to the incident direction of the ion beam is 
critical for the resulting profile.
For polycrystalline targets, we must additionally be concerned 
with the typical size of the crystalline grains relative to the depth to 
which an ion travels as well as the width of the ion beam on the surface 
of the target. If the beam radius is small compared to the grain size, 
we may consider that only one grain orientation is relevant for the 
incoming ion. In this case one needs to have knowledge of the grain
orientations. If the depth of the traveling ion is 
larger than the typical grain size, then one needs to insert characteristic 
disruptions in the target material orientation during the simulation of each
individual ion track. However, in the $\rm UO_2$ SIMS 
experiments \cite{saidy08}, it is reported that an
ion radiation depth-profile (of hundreds of keV) extends to less than a micron,
while the typical grain size of the polycrystalline samples is about 5 $\mu$m.
The applied beam size $150\times 150$ $\rm \mu m^2$ is, conversely, much 
larger than the grain size. Thus, the physical situation is as sketched 
in Fig. \ref{fig:poly}, where each ion-track simulation can
be represented as an implant into a crystalline material, but with 
the crystal orientation randomly selected for that particular simulation
(note that we are making the simplifying assumption that the sample is not
textured and does not possess preferred relative grain orientations). 
The ensemble of such simulations is then assumed to be representative 
of the overall 
depth profile. We note that more sophisticated target material 
representations, in which grain boundaries are included in a simulation, 
could be adopted, but we would then need to know more about the characteristics
of grain sizes from orientation imaging microscopy for instance. 
Given that this information is not provided for the 
experiments of interest, and that the ion range is significantly shorter 
than the reported typical grain size, we have conducted the simulations 
with the least amount of complexity.

Recent iodine implant experiments with SIMS analyses \cite{saidy08} include
implant energies between 100 keV and 800 keV, and SIMS conducted after
post-implant diffusion has been allowed. We are not concerned with
diffusion properties in this work, and we will therefore focus exclusively 
on the ''as-implanted'' SIMS results. These experiments can be directly 
simulated with the present approach.

\begin{figure}
   \centering
   \includegraphics[clip, scale=0.45, angle=0]{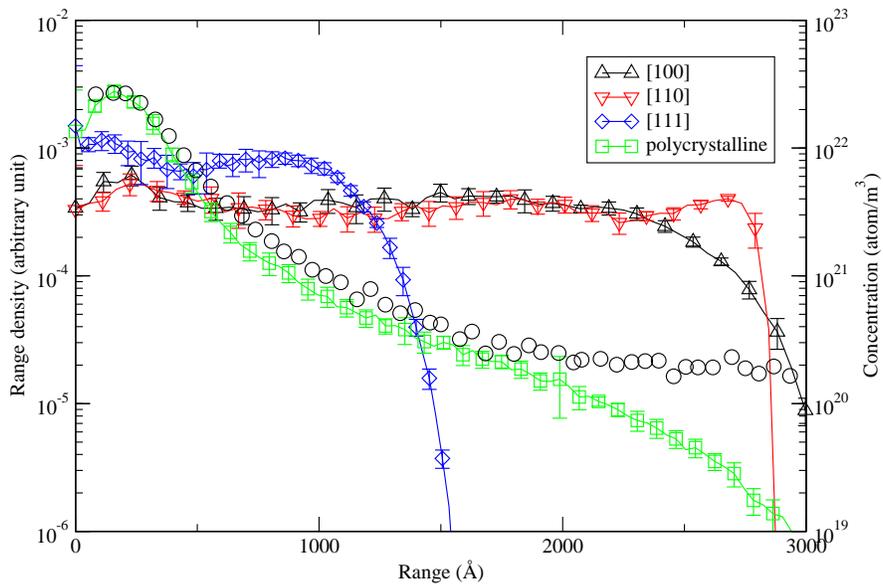}
\caption{Range density results of REED-MD for iodine 100 keV on $\rm UO_2$. 
Circular marks are the concentration data of iodine with polycrystalline  
$\rm UO_2$, sampled from Fig. 2 of Ref. \cite{saidy08}.}
\label{fig:i_100k}
\end{figure}
First, for 100 keV iodine, we have conducted simulations of implant depth 
profiles for both single and polycrystalline targets, where the single 
crystalline simulations have been conducted such that the implant directions 
are aligned with the characteristic UO$_2$ fluorite crystal 
orientations [100], [110], and [111]. The results, obtained at 300 K, 
are summarized in Fig. \ref{fig:i_100k}, where the line-connected
markers with error bars represent the four different simulation results. 
We observe the clear distinctions between the different orientations, 
and we confirm that the deepest implants at this energy correspond to the
orientations with the widest channels. It is noteworthy that the implant 
profiles for the polycrystalline sample are also distinct from the implants 
along the characteristic crystal orientations. This distribution is the
one exhibiting the most initial atomic collisions, signified by the 
large short range peak in the distribution, yet with more channeling ions 
than what is found for the small channel-area direction [111]. This
is consistent with a random orientation, since the initial peak represents 
all the off-axis implant events, while the few deep results are those where 
the implant direction has been aligned with one of the two large 
channel-area directions. The corresponding experimental data from Fig. 2 
of Ref. \cite{saidy08}, which is the result of measurements on polycrystalline
target materials, is included as circular markers. The agreement 
between our orientationally averaged simulation results and the 
experimental data for the polycrystalline target materials is 
near-perfect for ranges short of 0.2 $\mu$m, which accounts for about 
two orders of magnitude in the density. After this distance, 
the experimental density is nearly flat, whereas the simulated
profile continues to decay. This discrepancy can be due to the 
numerical representation of the polycrystalline target material. 
One may imagine that rapid diffusion may occur along grain boundaries; 
an effect that is not possible to model our simulations. However, 
one may also speculate that the experimental density beyond 0.2 $\mu$m
could reflect instrument resolution in the SIMS measurements. In any case, 
we consider 
the agreement convincing, and emphasize that any significant
discrepancy occurs at densities two orders of magnitude below the peak value.

With this information, it is curious to observe the comparable data for 
single-crystalline target materials shown in Ref \cite{saidy08}.
This density profile looks very similar to that 
of the polycrystalline sample, and nothing like the simulated on-axis
density profiles we report in Fig. 2. We cannot explain this discrepancy 
with any certainty, since we do not know the details of the materials 
processing or specific orientation of the surface.

\begin{figure}
   \centering
   \includegraphics[clip, scale=0.4, angle=0]{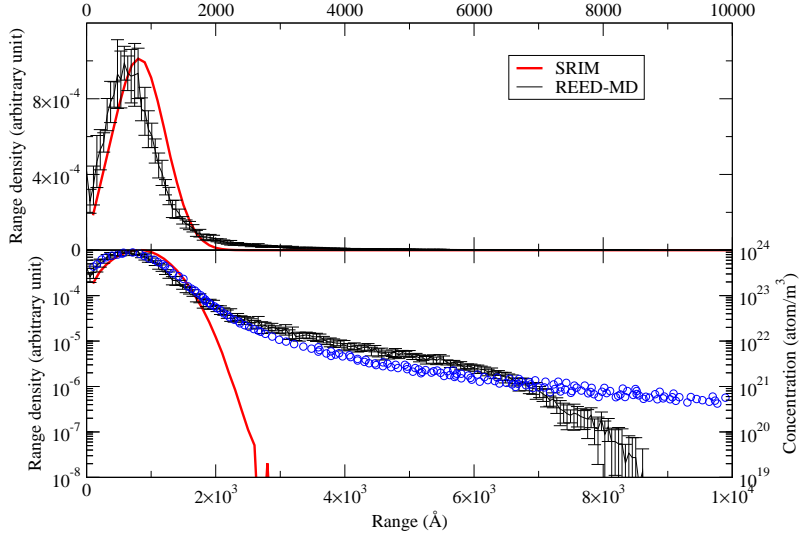}
\caption{Range density results of REED-MD and SRIM for iodine 440 keV on 
polycrystalline $\rm UO_2$. The top plot shows the results of SRIM and 
REED-MD on a linear scale while the bottom plot is on a log-scale. 
Circles on the bottom plot are  iodine concentration data, sampled from 
Fig. 12 of  Ref. \cite{saidy08}.} 
\label{fig:i_440k}
\end{figure}

Next, we simulated the reported \cite{saidy08} 440 keV iodine implant into
polycrystalline $\rm UO_2$. In addition to comparing the as-implanted SIMS
profile to our REED-MD simulations, we have also included results of
SRIM (Stopping and Range of Ions in Matter) \cite{SRIM} 
simulations in order to display the significance of 
channeling and structured target interactions
guiding the moving ion. The results are displayed in Fig. \ref{fig:i_440k},
where the top plot shows the comparison between REED-MD and SRIM on a 
linear scale of density. It is obvious that the primary, collisional 
peak in the distribution is well characterized by the BCA.
There are only 
minor differences in the peak location, with the REED-MD results showing 
a slightly shorter depth. However, when looking at the data on a logarithmic 
scale of density, the lower part of Fig. 3,
we see how the channeling tail of the profile is entirely absent in BCA. 
While this tail does not contain a large fraction of the implanted ions, 
it does show the extent of the implant range being several times larger than
that simulated with BCA. Also on the lower plot of Fig. 3 is shown the 
experimental data as open circular markers. We again observe that the 
experiments support the profile obtained by REED-MD throughout three orders
of magnitude of density, and we observe that the location of the primary 
peak in the distribution is also in agreement. As was the case for the 
results of 100 keV implants, discussed above, the longer tail of
the experimental distribution may be due to either the noise level in 
the SIMS profile, high dose effects, or some fast diffusion along 
grain boundaries during the time frame of implantation.

\begin{figure}
    \centering
    \includegraphics[clip, scale=0.4, angle=0]{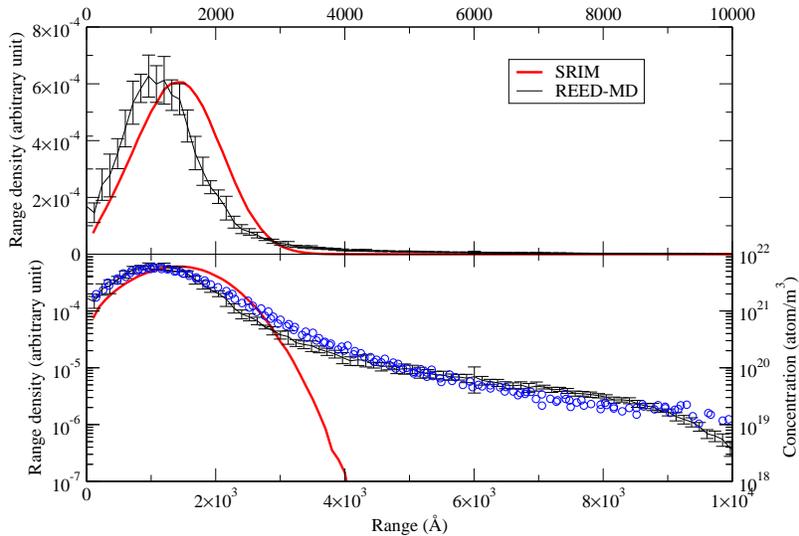}
\caption{Range density results of REED-MD and SRIM for iodine 800 keV on 
polycrystalline $\rm UO_2$. The top plot shows the results of SRIM and 
REED-MD on a linear scale while the bottom plot is on a log-scale.  
Circles on the bottom plot are  iodine concentration data, sampled from 
Fig. 1 of Ref. \cite{saidy08}.}
\label{fig:i_800k}
\end{figure}

Finally, we compare REED-MD with SRIM and the reported SIMS data for 800 keV 
iodine into polycrystalline samples. The results are shown in 
Fig. \ref{fig:i_800k}, and the observations are very similar to those made 
for the 440 keV implants discussed above. SRIM and REED-MD give
similar primary peaks of the distribution, but the BCA cannot adequately 
account for the extended tail of the distribution. 
The REED-MD result is validated by the experiments, both in the
location of the primary peak and the extent of the long tail of 
the distribution, which agree for the entire range of data that is 
available. 

We tested $\rm UO_2$ nuclear fuel materials implanted with high energy iodine 
and found that our method can predict the range profile very well, especially
for the poly crystal.

\section{Discussion and Conclusion}
We have provided direct comparisons between published experimental SIMS
data for high-energy iodine implants into polycrystalline uranium-dioxide 
samples and comparable REED-MD simulations.
The comparisons for all three reported energies are very
encouraging and provide validation of the modeling approach as a predictive 
tool for studying ion ranges in oxide fuels. The observed discrepancies for 
polycrystalline samples are minor, and only occur at the far range of the 
distributions where many plausible origins of discrepancies can be imagined. 
These include fast diffusion along grain boundaries during the time of 
implantation, and the possible instrument resolution limits in the SIMS
measurements.

Our simulations are of specific importance for modeling the 
results of secondary ion mass spectrometry measurements.  
In a more general sense though, the present investigation shows that 
accurate implantation profiles may be simulated.  This ability opens the 
possibility of using REED-MD simulations to aid in evaluating new fuel 
materials. Further validation of the modeling parameters can benefit from
detailed experiments of well characterized on-axis implantations, since 
Fig. 2 illustrates the sensitivity  of the range profiles as a function of
implant direction.

\section*{Acknowledgments}
This work was supported by the US Department of Energy 
Nuclear Energy Research Initiative for Consortia (NERI-C) contract number
DR-FG07-071D14893, by the Materials Design Institute at Los Alamos National
Laboratory (subcontract number 75782-001-09), and by Los Alamos National
Laboratory contract No. 75287-001-10.

\section*{References}

%\bibliographystyle{elsart-num}
%\bibliography{bib_reed}

\end{document}